# Overcurrent Protection for Gyrotrons in EAST ECRH System

Weiye Xu [1*], Handong Xu [1], Jianqiang Feng [1**], Yong Yang [1], Huaichuan Hu [1], Fukun Liu [1], Yunying Tang [1]

[1] Institute of Plasma Physics, Chinese Academy of Sciences, 350 Shushanhu Rd., Hefei, Anhui, China
[*] xuweiye@ipp.cas.cn
[**] jqfeng@ipp.cas.cn

**Abstract:** In this paper, an overcurrent protection system is designed to ensure the safety of the gyrotrons in the EAST ECRH system. Two overcurrent protection systems were established, a fast one and a slow one. The fast one uses the current transformers as the current sensors. The models of the current transformers and the superconducting magnet were built to analyze the effect of the environmental magnetic field on the current transformers using FI method. The analysis results show that the magnetic induction at the position near the current transformers must less than 0.002 T, i.e., the current transformers should be placed at a distance greater than 2.2 meters from the magnet center to ensure its normal work. The slow one uses the shunt to monitor the currents. An anti-fuse FPGA and a timer is used to realize the signal processing in the fast protection circuit and the slow protection circuit respectively. The response time of the fast protection circuit is less than 100 ns, and the response time of the slow protection circuit is less than 31 μs.

## 1. Introduction

Tokamak is one of the most promising methods to achieve the magnetic confinement nuclear fusion. A 140GHz/4MW/100s ECRH (electron cyclotron resonance heating) system for EAST (Experimental Advanced Superconducting Tokamak) is being built in ASIPP (Institute of Plasma Physics, Chinese Academy of Sciences)[1]. The ECRH system includes four gyrotrons[2] which are capable of producing 900~1000 kW of RF output power[3].

In EAST ECRH system, the overcurrent protection is one of the most important parts of the supervisory control and protection system. The gyrotrons are electric vacuum devices, they are very fragile, and must be operate very carefully. In the case of internal breakdown or bad electron beam focusing, the cathode current, the anode current will exceed the normal value. In that case, the high voltage power supplies of the gyrotrons must be shut off immediately to protect the gyrotrons from broken. Therefore, a reliable control and protection system, especially the overcurrent protection system must be established to ensure the safety of the gyrotrons.

In this paper, an overcurrent system is designed. Some EMC problems for overcurrent protection are discussed. The details are discussed in the following sections. In Section 2, the architecture of the supervisory control and protection system are given. In Section 3, the design and the analysis of the overcurrent protection system is given. Then, in Section 4, we give the conclusions.

## 2. Supervisory Control and Protection System for ECRH

In order to ensure the safety and the reliable operation of the gyrotrons, a supervisory control and protection system has been developed. It can be decomposed into nine subsystems, namely: central control system, state monitoring and the interlock system (use PLC), video monitoring system, overcurrent protection system, arc protection system, polarization control system, data acquisition system, power measurement and control system, data management and publishing system. Where the central control system is used to control the timing of the gyrotron operation; the state monitoring and the interlock system implements the interlock of all the subsystems; the video monitoring system is used for video recording of the heating device room where the gyrotrons are placed in order to monitor the status of the heating device room; the overcurrent protection system is one of the most important protection systems to protect the gyrotrons from broken, and it is the focus of this paper. The arc protection system monitors the arc signals of the MOU, the output window, and the miter bends to prevent these devices from being damaged by the arc; the polarization control system is used to control the polarization of the millimeter wave so that the electron cyclotron wave can be mostly absorbed by the plasma; The data acquisition system is used to acquire and record the relevant data of the gyrotrons. The power measurement and control system is responsible for measuring and control the output power of the gyrotrons. The data management and distribution system is used to store the relevant data and provide a convenient way to access data.

The architecture of the supervisory control and protection system are shown in Fig. 1. Whenever the current exceeds the threshold value, the overcurrent protection system will send the protect signals to the cathode power supply, the anode power supply, the PLC for interlock, and the center control system to shut off the power supplies and interlock the whole system. The cathode power supply and the anode power supply can be shut off within 10 μs[4].

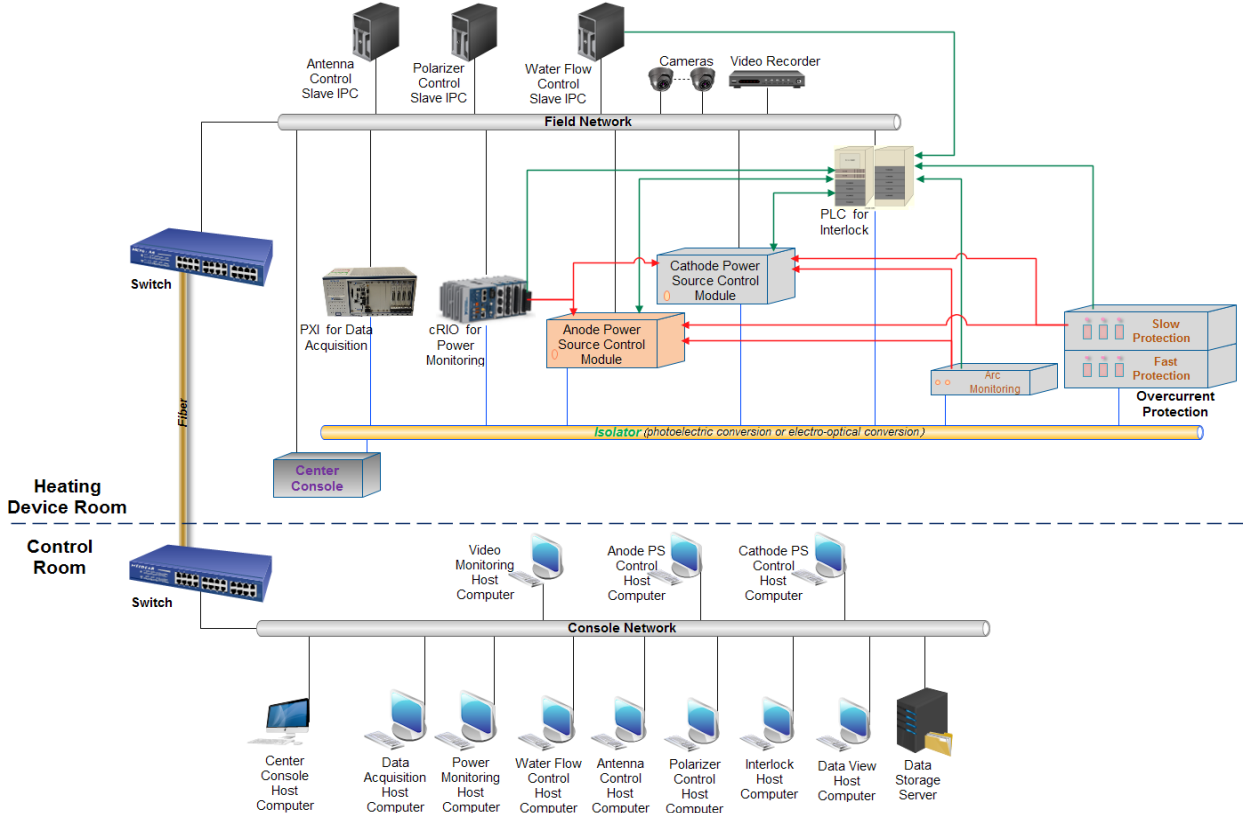

**Fig. 1.** Architecture of the supervisory control and protection system. The green cables indicate the connections between the devices and the PLC device, the red cables represent the protection signals from the devices to the high voltage power supply, the black cables indicate the network cables, and the blue cable indicates the connections between the devices and the center console.

### 3. Overcurrent Protection System for Gyrotrons

In the case of internal breakdown or bad electron beam focusing, the arc may happen in the gyrotrons while the cathode current, the anode current will exceed the normal value. It is very dangerous for gyrotrons because they may be damaged by arc. So, the high voltage power supplies of the gyrotrons must be shut off immediately whenever the current exceeds the threshold value to protect the gyrotrons from broken. Therefore, a reliable overcurrent protection system has to be established to ensure the safety of the gyrotrons.

According to the safety requirements of the gyrotrons, the overcurrent protection must ensure that the high-voltage output is cut off within 10 μs when the overcurrent happens. The permissible arc energy for the gyrotron gun is 10 J. Two sets of overcurrent protection systems who work in parallel were designed. The current detection principle is shown in Fig. 2.

A shunt whose resistance value is 500 mΩ is used to measure the anode current for slow overcurrent protection. A shunt whose resistance value is 1 mΩ is used to measure the cathode current for slow overcurrent protection. Four current transformers (two current transformers whose model is Pearson 2100 are used to sense the cathode current, and two current transformers whose model is Pearson 110 are used to sense the anode current) are used for fast overcurrent protection. The two sets of protection systems will be described separately in the following sections.

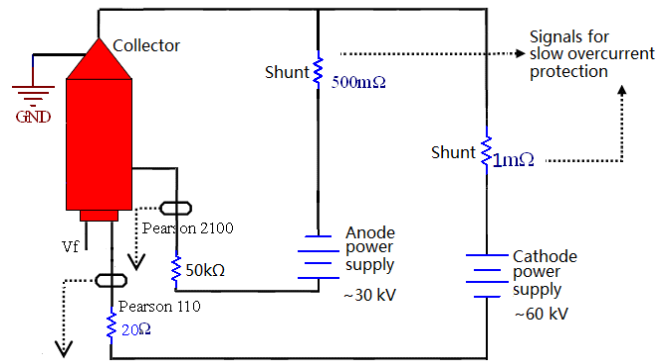

**Fig. 2.** The principle of current detection. Two shunts are used to measure the cathode current and the anode current for slow overcurrent protection. Two current transformers are used for fast overcurrent protection.

#### 3.1. Fast Overcurrent Protection

The functional block diagram of the fast protection system is shown in Fig. 3. The high-speed current transformers are used in the fast protection system to monitor the anode current and the cathode current. The cathode current and anode current are sensed by Pearson 110 and Pearson 2100 respectively[5]. The output signals of the current transformers are sent to the rapid protection circuit through attenuators (×1 attenuation for anode current, ×10 attenuation for cathode current). When the signal amplitude exceeds the preset threshold, the rapid protection circuit will send out the protection signals to shut off the high voltage power supplies. In order to achieve the isolation between subsystems, reduce line interference, and improve electromagnetic compatibility, the protection signals are converted to optical signals which can be transmitted by fiber. The fast protection circuit has the functions such as self-test, latch, reset, and so on. The threshold can be set remotely by the control room host computer through the communication interface.

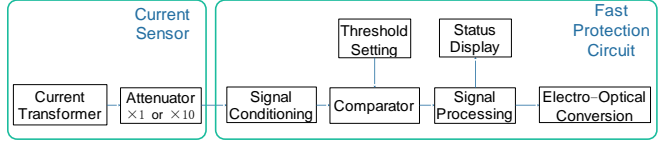

*Fig. 3. The functional block diagram of the fast protection system.*

The fast switching diode is used to clamp the input voltage signal as the signal conditioning circuit. The response time of the comparator is 7ns and the output signal is TTL level. The signal processing circuit uses an anti-fuse FPGA. The FPGA can latch the comparator and can control the latch time. During the period of the comparator latch, the comparator will always output a high level signal. All of the input and output signals to FPGA are isolated by the high-speed optocoupler with a typical response time of 50 ns. The FPGA is programmed with a watchdog function. Electro-optical conversion circuit, i.e. the optical drive circuit is implemented by using a fiber-optic transmitter. It can drive optical transmission distance of 1.25 km in the ideal case when the drive current is 30 mA. Under normal conditions, there is light output, and if there is no light, it means that the protection is occurring.

The PCB of the fast protection circuit uses four-layers. The top and bottom layer is the signal layers. The middle two layers is the power layer and ground layer respectively. The power plane was divided into two parts: 5V digital power supply and 5V analog power supply. The ground floor is also divided into two parts: digital ground and analog ground.

The fast protection circuit has been tested. The test results of the input–output logic is shown in Fig. 4 (a) and the test results of the response time is shown in Fig. 4 (b). In Fig. 4, the signal 3 is the input signal and the signal 4 is the protection output signal. As we can see, when the input voltage exceeds the threshold, the fast protection circuit will output high-level voltage; when the input voltage turns to low-level, the fast protection circuit will turn the output voltage to low-level. The response time is less than 100 ns.

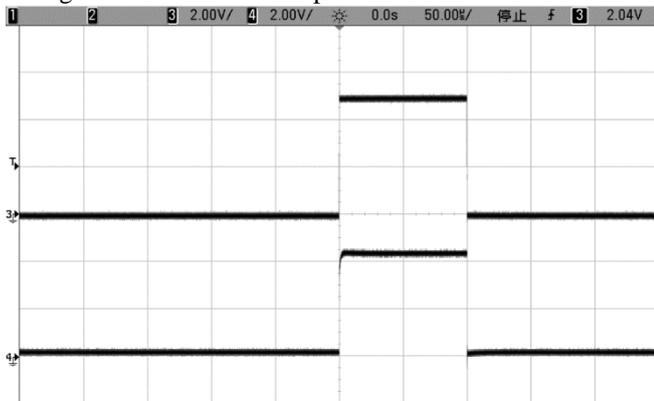
*a*

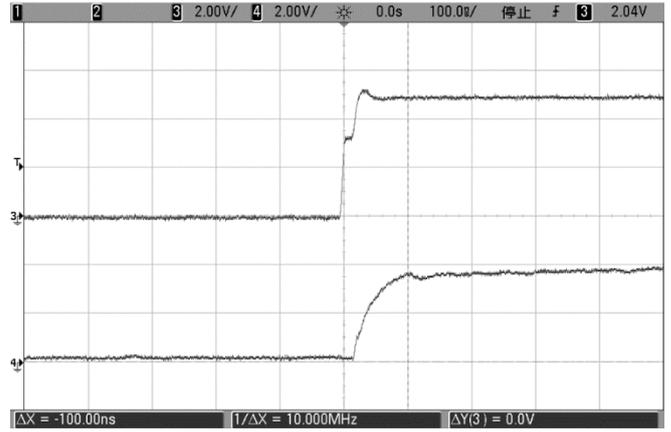
*b*

*Fig. 4. The test results of the fast protection circuit.*
*(a)* The input-output logic of the fast protection circuit. *(b)* The response time of the fast protection circuit.

As shown in Fig. 3. The high-speed current transformers are used in the fast protection system to monitor the anode current and the cathode current. The cathode current and anode current are sensed by Pearson 110 and Pearson 2100 respectively. Both the current transformers are single shielded. The transformers have insulated mounting brackets and the only grounding is through the cable braid.

The parameters of the Pearson current transformer we used (in the case of one turn) is shown in table 1. The useable rise time of the current transformers is 20 ns. If the rise time (the interval from 10% to 90% of the transition) of the input current pulse is less than the useable rise time, there will be an overshoot or ringing, and the amplitude of the overshoot or ring will be more than 10% of the transition.

The principle of measuring current using the current transformer is shown in Fig. 5. The secondary winding is wound on the core. The number of secondary windings is 50 for Pearson 2100, and the number of secondary windings is 500 for Pearson 110. The value of the output resistor R is 50 Ω for both two kinds of transformers. Because of the use of ferromagnetic core materials, and the current transformers are placed close to the superconducting magnet for gyrotron. The ferromagnetic cores can become saturated by the magnetic field. We analyze whether the magnetic field influences the normal use of current transformers in the followings.

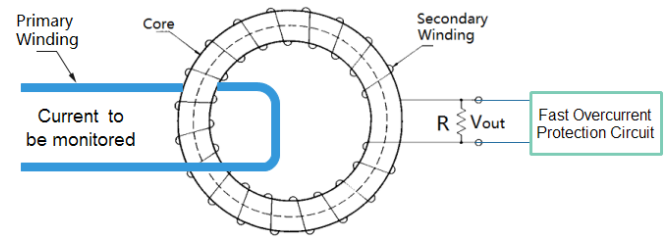

*Fig. 5. The schematic diagram of measuring current with the current transformer. The primary winding is in one turn.*

**Table 1** The parameters of the Pearson current transformer we used (in the case of one turn).

|  | Pearson 110 | Person 2100 |
| --- | --- | --- |
| Sensitivity [Volt/Ampere +1/-0%] | 0.1 | 1 |
| Output resistance [Ohms] | 50 | 50 |
| Maximum peak current [Amperes] | 5000 | 500 |
| Maximum rms current [Amperes] | 65 | 7.5 |
| Maximum DC current [Amperes] | 0.78 | 0.78 |
| Droop rate | 0.8% /ms | 0.08% /μs |
| Useable rise time [nanoseconds] | 20 | 20 |
| Current time product [Ampere-second max] | 0.5 | 0.005 |
| Low frequency 3dB cut-off (approximate) [Hz] | 1 | 125 |
| High frequency 3dB cut-off (approximate) [MHz] | 20 | 20 |
| I/f figure [peak Amperes/Hz] | 1.5 | 0.017 |
| Output connector | BNC (UG-290A/U) | BNC (UG-290A/U) |
| Operating temperature [℃] | 0 to 65 | 0 to 65 |
| Weight [ounces] | 22 | 21 |
| Hole diameter [inches] | 2.1 | 4.0 |
| Outside diameter [inches] | 2.1 | 4.0 |

As shown in table 1, the maximum DC current is 0.78A for both of the model 110 and the model 2100. The core in model 110 and the core in model 2100 are made out of the same material, that is supermalloy, a magnetically soft material. Unfortunately, we do not know the magnetization curve of the core. But we can estimate the magnetization curve of the core according to the parameters of the Pearson current transformers which are shown in table 1.

Firstly, we should need to model the current transformers. According to the parameters of the Pearson current transformers which are shown in table 1, we know that the size of model 110 and model 2100 are exactly the same. The only structure difference between model 110 and model 2100 is the number of secondary windings. The number of secondary windings can be calculated by this formula,

$$N = \frac{R}{S} \quad (1)$$

where $S$ is sensitivity; $R$ is the value of the output resistor which is 50 Ω for both two kinds of transformers. So, the number of secondary windings is 50 for Pearson 2100, and the number of secondary windings is 500 for Pearson 110.

The model we built are shown in Fig. 6. The primary winding is a solid coil with one turn. The secondary winding is a stranded coil whose turn is 50 for model 2100, and is 500 for model 110. The core is a nonlinear material. The FI[6, 7] (finite integration) method is used for the electromagnetic simulation. The S-Parameters obtained in the electromagnetic simulations is introduced into the circuit file to run the EM/circuit co-simulation. The co-simulation circuit model is shown in Fig. 6 (b).

Set the primary current as 0.78A, the simulation result shown in Fig. 7 (a) shows that the magnetic field strength on the core is H≈4 A/m. The relationship between magnetic induction B and magnetic field strength H is,

$$B = \mu H = \mu_r \mu_0 H \quad (2)$$

where,

$$\mu_0 = 4\pi \times 10^{-7} \text{ T} \cdot \text{m/A} \quad (3)$$

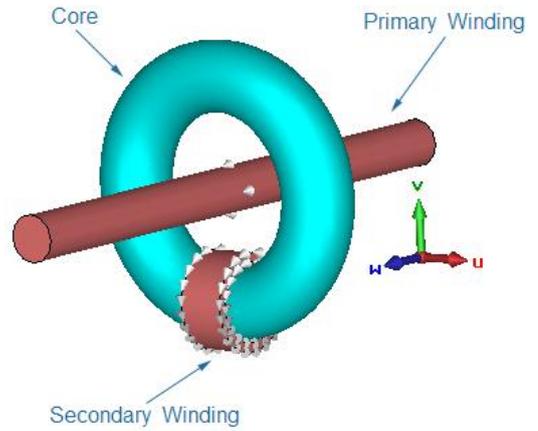

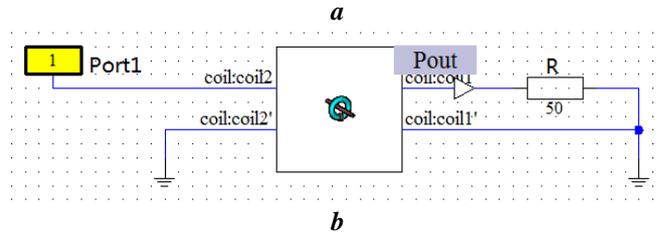

*b*

***Fig. 6.*** *The simulation model of the current transformer we built.*
*(a)* The model of the current transformer. *(b)* The schematic of the circuit model. Coil 2 represents the primary winding, and coil 1 represents the secondary winding.

Bringing H≈4 and μ0 into (2), we can got the relationship between the maximum relative permeability $\mu_{rmax}$ and the saturation magnetic induction $B_{max}$,

$$\mu_{rmax} \approx 198943.68 B_{max} \quad (4)$$

Because this formula is got by assuming the relative permeability $\mu_r$ is a constant which is equal to the maximum relative permeability $\mu_{rmax}$, the calculated saturation magnetic induction $B_{max}$ is larger than the real saturation magnetic induction $B_s$.

Because the core in model 110 and the core in model 2100 are made out of the same material. We estimated the magnetization curve of the core by analyzing the droop rate of model 2100. The droop rate is the downward slope of the top of the output voltage pulse resulting from a flattop current input pulse. Because the current monitor cannot pass DC, whenever the output voltage is non-zero and the current is constant, as during the flat top of the pulse, the voltage decays toward zero exponentially. We found that the greatest impact on the drop rate is the relative permeability. When the relative permeability is set as 25k, the drop rate of the model 2100 is approximately equal to 0.08% /μs. Using formula (4), we got the saturation magnetic induction $B_{max} \approx 0.126$ T. The optimized saturation magnetic induction is about 0.07 T. The optimized relative permeability curve and the magnetization curve is show in Fig. 7 (b) and Fig. 7 (c) respectively. When a square wave current signal excited at Port1 shown in Fig. 6 (b), which is shown in Fig. 7 (d), the output voltage at point Pout is shown in Fig. 7 (e). The drop rate is about 0.073% /μs which is very close to the real drop rate. When the dynamic current is 1 A, the dynamic output voltage is 1 V, which is consistent with the actual results.

If a magnetic induction of 0.005 T is applied in the x direction, it can be seen that the magnetic induction on the surface of the magnet reaches 0.07 T, which is shown in Fig. 7 (f). When a sine current signal of 1 A peak value excited at Port1, the output peak voltage is just about 0.4 V (shown in Fig. 7 (h)), which is less than the normal peak value 1 V. That is, the environmental magnetic field leads to the magnetic saturation of the core, the current transformer can not work normally. If a magnetic induction of 0.0025 T is applied in the x direction, it can be seen that the magnetic induction on the surface of the magnet reaches 0.05 T (shown in Fig. 7 (g)). When a sine current signal of 1 A peak value excited at Port1, the output peak voltage is about 1 V (shown in Fig. 7 (i)). If the magnetic field is applied in the y or z direction, the results are similar to that in the x direction. So, in order to ensure the normal operation of the current transformer, the current transformer should not be put in the environmental magnetic field whose magnetic induction greater than 0.002 T. For the model 110, the same material is used for the core, the simulation results are similar to the parameters shown in table 1. The effect of environmental magnetic field on model 110 is also similar to that on model 2100.

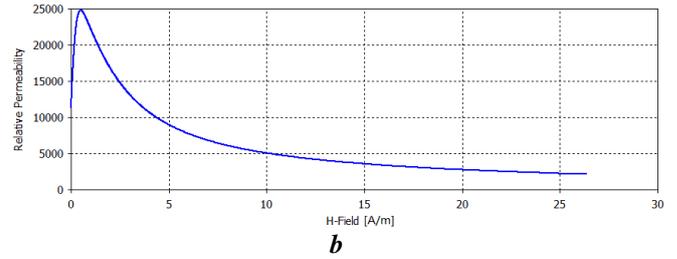

*b*

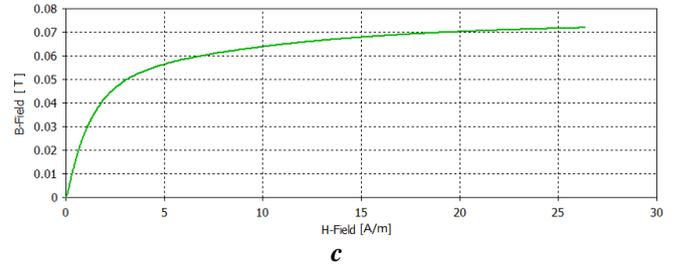

*c*

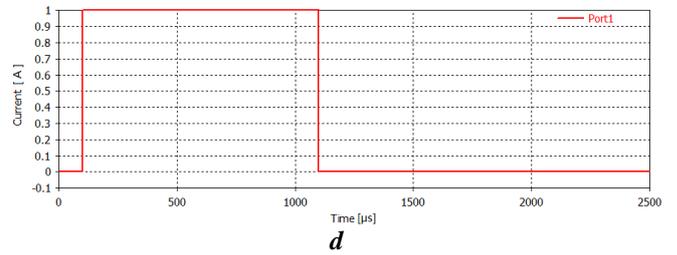

*d*

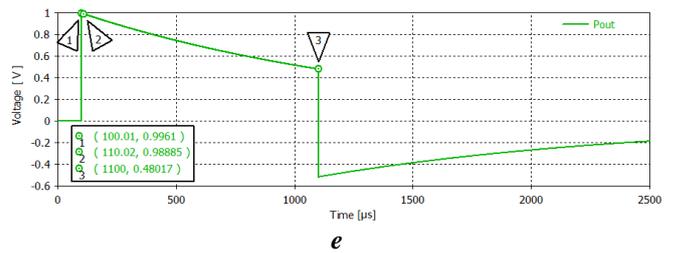

*e*

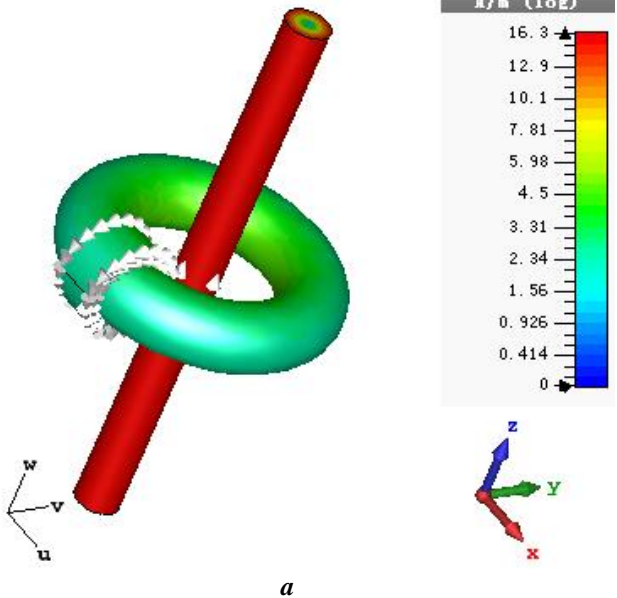

*a*

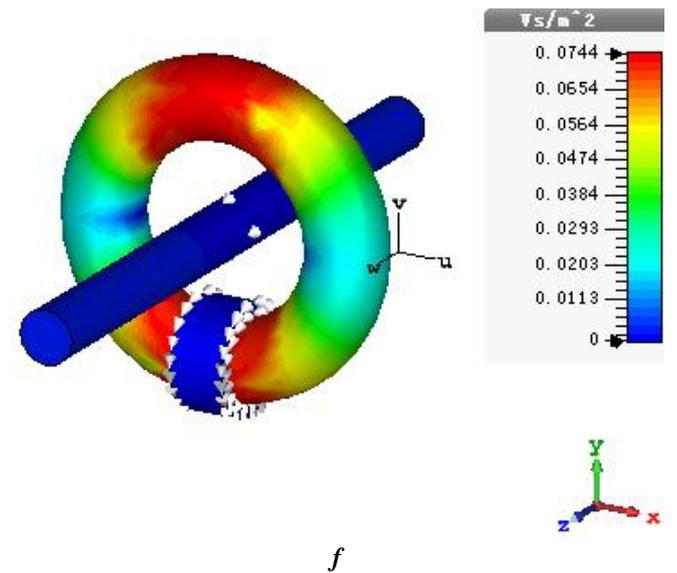

*f*

turns of the coil is 40000. The resistance and current of the superconducting coil is set as 0 Ω, 50.4 A, respectively.

The simulation results of the magnetic induction along with the distance from the magnet is shown in Fig. 9. According to the simulation result, we can see that the distance between the current transformers and the center of the magnet need to be greater than 2.2 m to insure the normal operation of the current transformers.

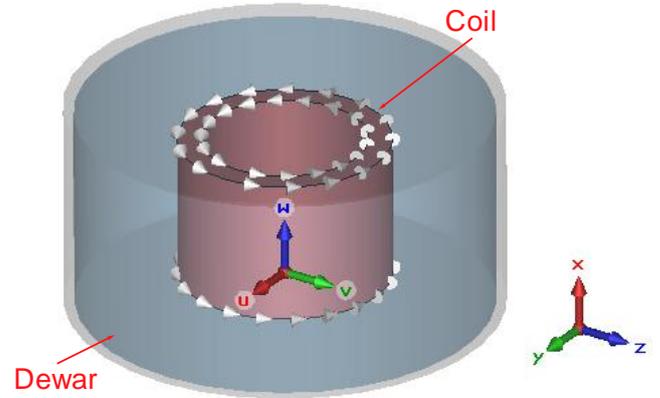

*Fig. 8. The simulation model of the superconducting magnet for gyrotron. The Dewar shell material is steel and the vacuum is filled inside the Dewar.*

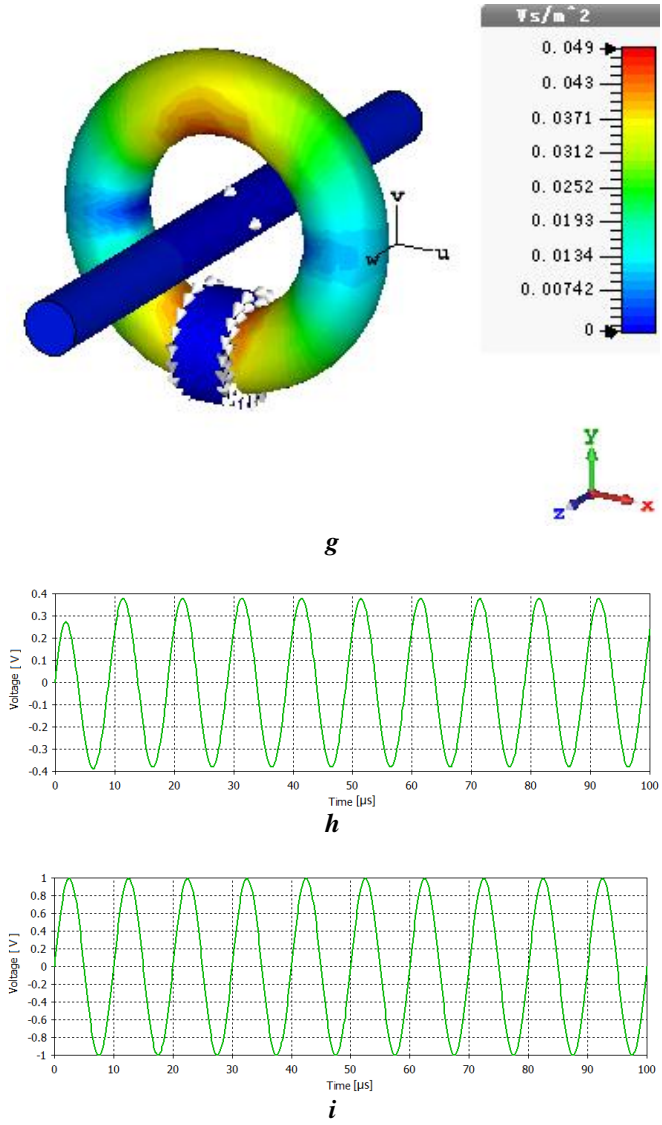

*Fig. 7. The simulation results of the current transformer.*
*(a)* The simulation result of the magnetic field strength when the primary current is 0.78 A. *(b).* The relative permeability of the core along with the magnetic field strength. *(c)* The magnetization curve (B-H curve) of the core. *(d)* The input current signal for transient analysis. *(e)* The simulation result of the drop rate. *(f)* The simulation result of the magnetic induction with a static magnetic field of 0.005 T applied in the x direction. *(g)* The simulation result of the magnetic induction with a static magnetic field of 0.0025 T applied in the x direction. *(h)* The output voltage when a sine current signal with 1 A peak value excited at Port1 with a magnetic induction of 0.005 T is applied in the x direction. *(i)* The output voltage when a sine current signal with 1 A peak value excited at Port1 with a magnetic induction of 0.0025 T is applied in the x direction.

In order to ensure that the environmental magnetic induction near the current transformer is smaller than 0.002 T, we should analyze the magnetic field generated by the superconducting magnet. The simplified model of the magnet for gyrotron is shown in Fig. 8. The inner diameter of the superconducting coil is 13.8938 cm, the outer diameter is 17.8587 cm, and the height is 0.247626 cm. The number of

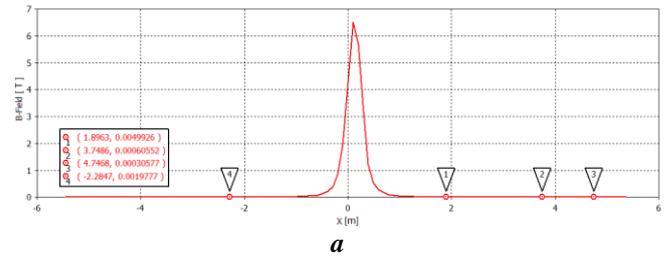

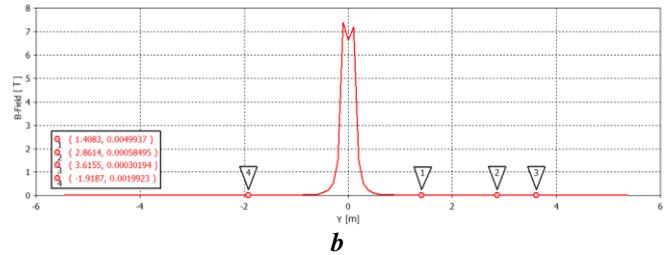

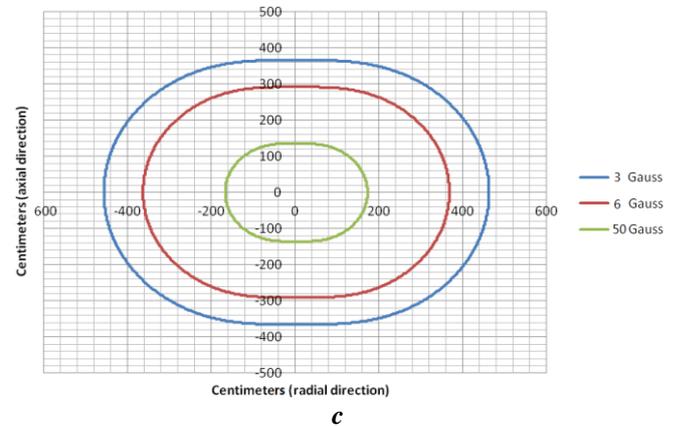

*Fig. 9. The simulation results of the superconducting magnet.*

*(a)* The magnetic induction along with the x direction (axial direction). In the simulation model, the axial center is at the position of x=0.123813 m. The magnetic induction is the largest at the center position. *(b)* The magnetic induction along with the y direction (radial direction). In the simulation model, the radial center is at the position of y=0 m. *(c)* The magnetic induction along with the distance from the center. For instance, at the point whose coordinate is (Radial, axial) = (140 cm, 180 cm), the magnetic induction is about 0.005 T (50 Gauss).

In the actual application, the current transformer is placed at a distance greater than 2.2 meters from the magnet. The fast overcurrent protector can work stably. Whenever the overcurrent happens, the protection signals will be send to the power supplies, the PLC, and the center controller to shut off the gyrotrons. Fig. 10 shows an example of the signals output by the current transformers when a gyrotron overcurrent happens.

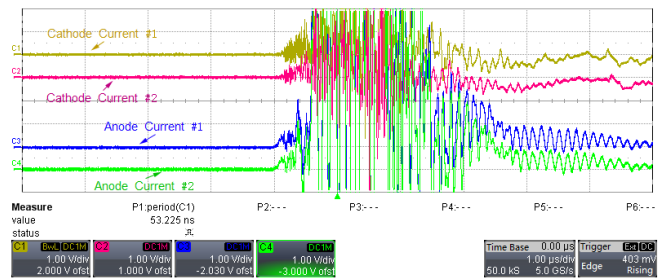

*Fig. 10. An example of the signals output by the current transformers when a gyrotron overcurrent happens. The cathode current is monitored by two model 110 current transformers with ×10 attenuators, so the cathode current is 100 A/div in this figure. The anode current is monitored by two model 2100 current transformers with no attenuator, so the anode current is 1 A/div in this figure.*

### 3.2. Slow Overcurrent Protection

In order to further ensure the safety of the gyrotrons. A slow overcurrent protection system is designed and established. In order to detect the steady current during the normal operation of the anode current and cathode current, the shunts (1 mΩ for cathode current, 500 mΩ for anode current) are used in this slow protection system. The block diagram is shown in Fig. 11.

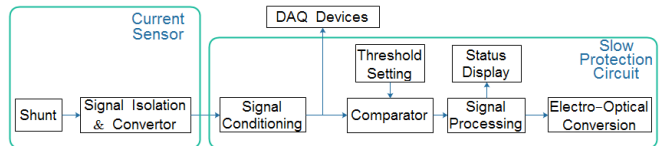

*Fig. 11. The functional block diagram of the slow protection system.*

The shunt is actually a resistor with a small resistance value that can withstand high power. The signal isolation & convertor can convert the 0 ~ 100 mV voltage signal to 4 ~ 20mA current signal.

The signal conditioning circuit of the slow protection circuit is a little different with that of the fast protection circuit. Besides a fast switching diode is used to clamp the input voltage signal, a 250 Ω resistor is used to convert the 4 ~ 20 mA current signal to 0 ~ 5 V voltage signal. This voltage signal is transmitted to the comparator, and to the DAQ (data acquisition) devices such as PXI DAQ cards[8], and the oscilloscope in parallel. The response time of the comparator is 200 ns and the output signal is TTL level. The signal processing circuit is implemented by the 555 timer. Whenever an overcurrent happens, the timer will be triggered to output a high-level signal for a period of time $\tau=1.1\ C_t R_t$. For this circuit, the duration $\tau$ is from 0.22 s to 11.22 s. The electro-optical conversion circuit of the slow protection is exactly the same as that of the fast protection.

We tested the signal isolation & conversion module and the slow protection circuit. The test results show that the response time of the signal isolation & convertor is less than 30 μs, and the response time of the slow protection circuit is about 200 ns. The total response time of these circuit is less than 31 μs.

### 4. Conclusion

The overcurrent protection system is very important in the EAST ECRH system. It is used to monitor the cathode current and the anode current of the gyrotrons to ensure the safety of the gyrotrons. We have established two protection systems, i.e., the overcurrent fast protection system and the overcurrent slow protection system. The fast one uses the current transformers to monitor the cathode current and the anode current. The current transformer can be affected by the environmental magnet field because the core is made out of the magnetically soft material which may be led to the magnetic saturation by the magnetic field. So the current transformers can not work normally in an environmental magnetic field with bigger magnetic induction. We analyzed the effect of the environmental magnetic field on the current transformers using the FI method. The analysis results show that the magnetic induction at the position near the current transformers must less than 0.002 T, i.e., the current transformers should be placed at a distance greater than 2.2 meters from the magnet center to ensure its normal work. The slow one uses the shunt to monitor the currents. The shunt is actually a resistor with small resistance value and can withstand high power. The current is converted to the voltage by the shunt. The voltage signal is transmitted to the comparator and the DAQ devices in parallel.

The protection circuits of these two protection systems are similar. The mainly different between them is that the fast one use the anti-fuse FPGA to process the signals and the slow one use a timer to realize the signal processing. Using anti-fuse FPGA is more stable and easy to expand the functions of the protection system, but it is more expensive. The response time of the fast protection circuit is less than 100 ns, and the response time of the slow protection circuit is less than 31 μs.

The experimental shows that both two overcurrent protection systems can work normally and stably. The gyrotrons can always work securely under the protection of the overcurrent protection system.

### 5. Acknowledgments

This work was supported by the National Magnetic Confinement Fusion Science Program of China (Grant No. 2015GB103000). We greatly appreciate the experts from GA,